\documentclass[twocolumn,aps,prl,superscriptaddress,showpacs,preprintnumbers,amsmath,amssymb]{revtex4-1}
\usepackage{graphicx,color}
\usepackage{dcolumn}
\usepackage{colordvi} 
\usepackage{bm}

\usepackage{amsmath,amssymb,amsfonts}
\usepackage[]{color}

\begin{document}
\title{Universally Valid Error-Disturbance Relations in Continuous Measurements}
\author{Atsushi Nishizawa}
\email{anishi@caltech.edu}
\affiliation{Theoretical Astrophysics 350-17, California Institute of Technology, Pasadena, California 91125, USA}
\author{Yanbei Chen}
\affiliation{Theoretical Astrophysics 350-17, California Institute of Technology, Pasadena, California 91125, USA}
\date{\today}

\begin{abstract}
In quantum physics, measurement error and disturbance were first naively thought to be simply constrained by the Heisenberg uncertainty relation. Later, more rigorous analysis showed that the error and disturbance satisfy more subtle inequalities.  Several versions of universally valid error-disturbance relations (EDR) have already been obtained and experimentally verified in the regimes where naive applications of the Heisenberg uncertainty relation failed. 
However, these EDRs were formulated for discrete measurements. In this paper, we consider continuous measurement processes and obtain new EDR inequalities in the Fourier space: in terms of the power spectra of the system and probe variables. 
By applying our EDRs to a linear optomechanical system, we confirm that a
tradeoff relation between error and disturbance leads to the existence of an optimal strength of the disturbance in a joint measurement. Interestingly, even with this optimal case, the inequality of the new EDR is not saturated because of doublely existing standard quantum limits in the inequality. 

\end{abstract}

\pacs{03.65.Ta, 42.50.-p, 42.50.Lc}
\maketitle

{\it{Introduction}} --- The uncertainty principle is one of the most fundamental features of quantum physics; it prevents us from measuring non-commuting observables simultaneously with arbitrary accuracy. The original formulation by Heisenberg \cite{aHeisenberg:1927zz} was based on a thought experiment that measures the position of an electron with a $\gamma$-ray microscope. Nowadays the Heisenberg uncertainty relation (HUR) is known as
\begin{equation}
\epsilon_A \eta_B \gtrsim \left| C_{AB} \right| \;.
\label{eq1}
\end{equation}
where $C_{AB} \equiv i \langle [ \hat{A},\hat{B}] \rangle /2$, $\epsilon_A \equiv \epsilon (\hat{A})$ is the error in the measurement of the observable $\hat{A}$ of a quantum system and $\eta_B \equiv \eta (\hat{B})$ is the disturbance exerted back onto a system observable $\hat{B}$, which does not commute with $\hat{A}$. 

Long after the original proposal of the HUR, more rigorous analyses of the measurement process incorporating the measuring device, or {\it probe}, revealed that the HUR, when used naively as an error-disturbance relation (EDR), can be violated, e.g.~\cite{Ozawa:2002PhLA}. Ozawa proved a revised, universally valid EDR \cite{Ozawa:2003PhLA,Ozawa:2003PhRvA}:
\begin{equation}
\epsilon_A \eta_B + \epsilon_A \sigma_{B} + \sigma_{A} \eta_B \geq \left| C_{AB} \right| \;, 
\label{eq2}
\end{equation}
where $\sigma_A^2$ and $\sigma_B^2$ are variances of the observables $\hat{A}$ and $\hat{B}$, respectively. Soon after the proposal of the Ozawa's EDR inequality, it has been generalized to the error-tradeoff relation \cite{Ozawa:2004PhLA,Hall:2004PhRvA}, in which the disturbance $\eta_B$ is replaced with $\epsilon_B$ and the inequality is symmetrized about the variables $\hat{A}$ and $\hat{B}$. The Ozawa's EDR can be tightened by using stronger geometrical inequalities \cite{Weston:2013PhRvL,Branciard:2013PNAS,Branciard:2014PhRvA}. The strongest one has been derived by Branciard \cite{Branciard:2013PNAS,Branciard:2014PhRvA} and is given by 
\begin{equation}
\epsilon_A^2 \sigma_{B}^2 + \sigma_{A}^2 \eta_B^2 + 2 \sqrt{\sigma_{A}^2\sigma_{B}^2-C_{AB}^2}\, \epsilon_A \eta_B \geq C_{AB}^2 \;.
\label{eq3}
\end{equation}
These EDRs have been experimentally tested with the three-state method \cite{Erhart:2012NatPh,Baek:2013NatSR} and with the weak-measurement method \cite{Rozema:2012PhRvL,Kaneda:2014PhRvL,Ringbauer:2014PhRvL} and found that the HUR is violated while the universal EDRs hold. 
The above EDRs were derived for measurements performed at instants of time, i.e., discrete measurements, and cannot be applied directly to continuous measurements: in a discrete measurement, one can define the error and disturbance by comparing quantities before and after the measurement. However, notions of "before and after" are irrelevant to a continuous measurement. 
In this paper, we redefine error and disturbance as the differences between "without and with the measurement" and extend the above EDRs to the continuous measurements.

{\it{Fundamental variables}} --- 
%
In general, we consider the quantum measurement of a {\it system} achieved via coupling to a device, or a {\it probe}; the system and probe have separate Hilbert spaces. The projection postulate of quantum mechanics is applied directly to the probe --- and acts indirectly on the system.  We suppose that a system observable $\hat{A}$ is the target of measurement, with the meter observable of a probe $\hat{M}$ that is projectively measured, and consider the disturbance back onto another system observable $\hat{B}$.

For a continuous measurement process, we define the error and disturbance operators with time-dependent Heisenberg operators as $\hat{N} (t) \equiv \hat{M} (t) - \hat{A}_0 (t)$ and $\hat{D} (t) \equiv \hat{B} (t) - \hat{B}_0 (t)$, respectively, where the subscript "0" stands for {\it free evolution}, for which the system and probe are decoupled. These are motivated by Ozawa's original definitions for discrete measurement \cite{Ozawa:2003PhLA,Ozawa:2003PhRvA}, but differ in their interpretations. Specifically, for a discrete measurement, $\hat{N}$ and $\hat{D}$  compare variables between before and after the measurement. In the continuous case, our $\hat{N}$ and $\hat{D}$ compare the difference of variables between with and without measurements. The error and disturbance in the measurement of $\hat{A}_0$ are 
\begin{align}
\epsilon_A^2 (t) &\equiv \langle \hat{N}^2 (t) \rangle = \langle \{ \hat{M}(t) - \hat{A}_0(t) \}^2 \rangle \;, 
\label{eq85a}  \\
\eta_B^2 (t) &\equiv \langle \hat{D}^2 (t) \rangle = \langle \{ \hat{B}(t) - \hat{B}_0(t) \}^2 \rangle \;. 
\end{align}
Here $\langle \cdots \rangle$ denotes ensemble average or time average for a stationary system. The variances of observables $\hat{A}_0$ and $\hat{B}_0$ are
\begin{align}
\sigma_A^2 (t) &\equiv \langle \{ \hat{A}_0(t) - \langle \hat{A}_0(t) \rangle \}^2 \rangle \;, \\
\sigma_B^2 (t) &\equiv \langle \{ \hat{B}_0(t) - \langle \hat{B}_0(t) \rangle \}^2 \rangle \;. 
\label{eq86}
\end{align}
These definitions of error, disturbance, and variances are general, but here we treat a stationary measurement, where any two-point correlation function depends only on time difference. Then the quantities in Eqs.~\eqref{eq85a}--\eqref{eq86} become independent of time.

{\it{Error-disturbance relation in Fourier space}} --- 
We first note that Eq.~(\ref{eq2}) continues to hold if we replace error and disturbance quantities, $\epsilon_{A}$ and $\eta_{B}$, by their time-dependent versions defined in Eqs.~\eqref{eq85a}--\eqref{eq86} --- and commutator $C_{AB}$ by the equal-time commutator $[\hat A_0(t), \hat B_0(t)]$. This is because the algebraic relations that had lead to the original inequality continue to hold in this situation. However, at a steady state, the inequality is the same for all $t$, and we only have one condition governing the total fluctuations.  In order to obtain further constraints, we shall assume the noise processes to be stationary, and are therefore characterized by their spectra.  
%

For the later use, we introduce a filter function $\Gamma(t)$, whose shape can be chosen arbitrarily; the filtered error and disturbance operators are
\begin{align}
\hat{{\cal{N}}} (t) &\equiv \int_{-\infty}^{t} dt^{\prime} \Gamma(t-t^{\prime}) \hat{N}(t^{\prime}) = \hat{{\cal{M}}} (t) - \hat{{\cal{A}}}_0 (t) \;, \nonumber \\
\hat{{\cal{D}}} (t) &\equiv \int_{-\infty}^{t} dt^{\prime} \Gamma(t-t^{\prime}) \hat{D}(t^{\prime}) =\hat{{\cal{B}}} (t) - \hat{{\cal{B}}}_0 (t) \;, \nonumber
\end{align}
while $\hat{{\cal{A}}}_0 (t)$, $\hat{{\cal{B}}}_0 (t)$, $\hat{{\cal{M}}} (t)$, and $\hat{{\cal{B}}} (t)$ are defined in a similar fashion using the same filter.  The new error $\varepsilon$ can be computed using
\begin{equation}
\epsilon_{\cal{A}}^2 \equiv \langle \hat{{\cal{N}}}^2(t) \rangle = \int_{0}^{\infty} \frac{d\Omega}{2\pi} \left| \Gamma (\Omega) \right|^2 S_{\epsilon} (\Omega) \;.
\label{eq89}
\end{equation}
Here $\Omega$ is angular frequency and we defined
\begin{equation}
\Gamma(\Omega) \equiv \int_{-\infty}^{\infty} d\tau\, \Theta(\tau) \Gamma(\tau)  e^{i\Omega \tau} \;, \nonumber
\end{equation}
with the Heaviside step function $\Theta (\tau)$ and the (one-sided) power spectral density $S_{\epsilon} \equiv S_{NN}$ of $\hat{N}$ \cite{PRLsupplement}. 
Note that for a stationary state the time dependence of $\epsilon_{\cal{A}}$ disappears. Similar to Eq.~(\ref{eq89}), $\eta_{\cal{B}}$, $\sigma_{\cal{A}_0}$, and $\sigma_{\cal{B}_0}$ are given by
\begin{align}
\eta_{\cal{B}}^2 &= \int_{0}^{\infty} \frac{d\Omega}{2\pi} \left| \Gamma (\Omega) \right|^2 S_{\eta} (\Omega) \;, 
\label{eq13} \\
\sigma_{\cal{A}}^2 &= \int_{0}^{\infty} \frac{d\Omega}{2\pi} \left| \Gamma (\Omega) \right|^2 S_{\sigma_{A}} (\Omega) \;, 
\label{eq95} \\
\sigma_{\cal{B}}^2 &= \int_{0}^{\infty} \frac{d\Omega}{2\pi} \left| \Gamma (\Omega) \right|^2 S_{\sigma_{B}} (\Omega) \;, 
\label{eq90}
\end{align}
where $S_{\eta}$, $S_{\sigma_{A}}$, and $S_{\sigma_{B}}$ are the power spectra of $\hat{D}$, $\hat{A}_0$, and $\hat{B}_0$, respectively.

Since the filtering does not change the Hilbert space of each observable and not affect the proof of Eq.~(\ref{eq2}), the EDR in Eq.~(\ref{eq2}) still holds for the filtered quantities:
\begin{equation}
\epsilon_{\cal{A}} \eta_{\cal{B}} + \epsilon_{\cal{A}} \sigma_{\cal{B}} + \sigma_{\cal{A}} \eta_{\cal{B}} \geq | C_{\cal{A}\cal{B}} | \;, \label{eq20}
\end{equation}
with the commutator, $C_{\cal{A}\cal{B}}\equiv i \langle [ \hat{{\cal{A}}}_0 (t), \hat{{\cal{B}}}_0 (t) ] \rangle/2$.

Substituting Eqs.~(\ref{eq89}), (\ref{eq13})-(\ref{eq90}) and choosing $\Gamma$ to be narrowbanded within a very narrow frequency bin near $\Omega$, the filtered version of the Ozawa's inequality in Eq.~(\ref{eq20}) can be written as \cite{PRLsupplement}
\begin{align}
\sqrt{S_{\epsilon} (\Omega) S_{\eta} (\Omega)} + \sqrt{S_{\epsilon} (\Omega) S_{\sigma_B} (\Omega)} &+ \sqrt{S_{\sigma_A} (\Omega) S_{\eta} (\Omega)} \nonumber \\
& \geq \hbar \left| \langle \hat{\chi}(\Omega) \rangle \right| \;.
\label{eq4}
\end{align}
where we have defined (using staionarity)  
\begin{equation}
[ \hat{A}_0(\Omega), \hat{B}^{\dagger}_0(\Omega^{\prime})] = 2\pi i \hbar \delta (\Omega-\Omega^{\prime}) \hat{\chi}(\Omega) \;. \label{eq66}
\end{equation}
Since Eq.~\eqref{eq4} applies to all frequencies, it generalizes Ozawa's EDR~\cite{PRLsupplement} into a spectral relation for our continuous measurement.  

\begin{figure}[t]
\centering
\includegraphics[width=65mm]{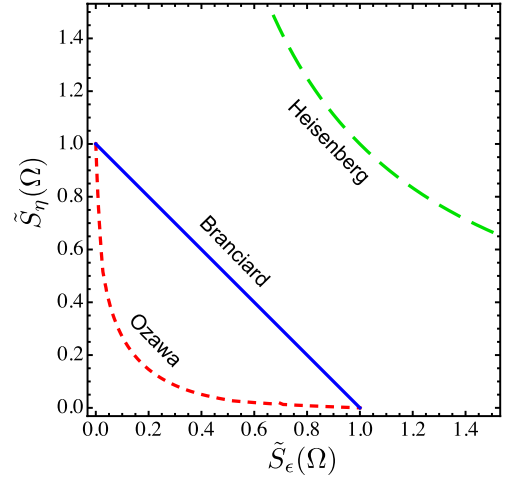}
\caption{Comparison of the EDR lower bounds. 
The curves indicate the locations at which the Ozawa (red short-dashed), Branciard (blue solid) and Heisenberg (green) inequalities are saturated. Here we assume $|\tilde{\chi}(\Omega)|=1$.
}
\label{fig6}
\end{figure}

Analogously, the generalization of the Branciard EDR to the continuous case reads \cite{PRLsupplement}:
\begin{align}
&S_{\epsilon} (\Omega)S_{\sigma_B} (\Omega) + S_{\sigma_A} (\Omega)S_{\eta} (\Omega) \nonumber \\
&+2 \sqrt{S_{\epsilon} (\Omega)S_{\eta} (\Omega) \left\{ S_{\sigma_A} (\Omega)S_{\sigma_B} (\Omega) - \hbar^2 \left| \langle \hat{\chi}(\Omega) \rangle \right|^2 \right\}} \nonumber \\
&\geq \hbar^2 \left| \langle \hat{\chi}(\Omega) \rangle \right|^2 \;. \label{eq82}
\end{align}

Defining $\tilde{S_{\epsilon}} \equiv S_{\epsilon}/S_{\sigma_A}$, $\tilde{S_{\eta}} \equiv S_{\eta}/S_{\sigma_B}$, and $\tilde{\chi} \equiv \hbar \langle \hat{\chi} \rangle/\sqrt{S_{\sigma_A}S_{\sigma_B}}$,
we obtain normalized Ozawa and Branciard inequalities:
\begin{align}
&\sqrt{\tilde{S}_{\epsilon} (\Omega) \tilde{S}_{\eta} (\Omega)} + \sqrt{\tilde{S}_{\epsilon} (\Omega)} + \sqrt{\tilde{S}_{\eta} (\Omega)}  \geq \left| \tilde{\chi}(\Omega) \right| \;,\label{eq84a} \\
&\tilde{S}_{\epsilon} (\Omega) + \tilde{S}_{\eta} (\Omega) +2 \sqrt{\tilde{S}_{\epsilon} (\Omega) \tilde{S}_{\eta} (\Omega) \left( 1 - \left| \tilde{\chi}(\Omega) \right|^2 \right)} \geq \left| \tilde{\chi}(\Omega) \right|^2 \;.
\label{eq83}
\end{align}

Let us briefly comment on the spectral expression of the Robertson's inequality \cite{Robertson:1929zz}, $\sigma_{A} \sigma_{B} \geq \frac{1}{2} |\langle [ \hat{A}_0,\hat{B}_0] \rangle|$, which is the basic uncertainty relation that holds for arbitrary system observables and has nothing to do with a measurement. Applying the filters in Eqs.~(\ref{eq95}) and (\ref{eq90}) (this is just a mathematical operation and not on actual data), we have the Robertson's inequality in the Fourier space,
\begin{equation}
\left| \tilde{\chi}(\Omega) \right| \leq 1 \;. \label{eq6}
\end{equation}
If the system saturates the Robertson's inequality, the third term in Eq.~(\ref{eq83}) vanishes. In other words, the Branciard inequality becomes the tightest when $\left| \tilde{\chi}(\Omega) \right| = 1$.

On the other hand, comparing Eqs.~(\ref{eq1}) and (\ref{eq2}), we find that the HUR is given by keeping only the first term on the LHS of Eq.~(\ref{eq84a}): 
\begin{equation}
\tilde{S}_{\epsilon} (\Omega) \tilde{S}_{\eta} (\Omega) \geq \left| \tilde{\chi}(\Omega) \right|^2 \;.
\label{eq101}
\end{equation}
As we shall illustrate later, there are the case where the Ozawa's inequality in Eq.~(\ref{eq84a}) is supported by the second and third terms on the LHS, thereby allowing the HUR to be violated. In Fig.~\ref{fig6}, three inequalities, Eqs.~(\ref{eq84a}), (\ref{eq83}), and (\ref{eq101}), are compared, setting $\left| \tilde{\chi}(\Omega) \right| = 1$.

{\it{Application to a linear optomechanical system}} --- We now apply our formalism to a linear optomechanical system, in which the position of a test mass is measured with an optical probe. Namely, we specify the system's observables $\hat{A}$ and $\hat{B}$ to be the position and momentum operators of the system, $\hat{x}$ and $\hat{p}$, and the meter variable $\hat{M}$ of the probe to be the phase quadrature of light $\hat{Q}$, whose unit is the same as $\hat{x}$. We assume that the interaction between the probe and the system is the von Neumann type, with Hamiltonian in the interaction picture given by $\hat{H}_I(t) = - \hat{x} _0(t) \hat{F}_{0}(t)$, where the subscript "0" denote free evolution and $\hat{F}_{0}$ is a generalized force.

For this linear system, the time evolution of any linear variable can be written in terms of its response to generalized forces. In Fourier space, if we define the response function
\begin{equation}
R_{XY}(\Omega) \equiv \frac{2}{\hbar} \int_{-\infty}^{\infty} d\tau\, \Theta(\tau) C_{XY}(\tau)  e^{i\Omega \tau} \;,
\label{eq50}
\end{equation}
which satisfies $R^{*}_{XY}(\Omega) = R_{XY}(-\Omega)$, then the dynamics of the system is described by \cite{PRLsupplement},
\begin{align}
\hat{x}(\Omega) &= \hat{x}_0(\Omega) + R_{xx}(\Omega) \hat{F}(\Omega)  \;, \label{eq27} \\
\hat{p}(\Omega) &= \hat{p}_0(\Omega) + R_{px}(\Omega) \hat{F}(\Omega) \;,  \\
\hat{Q}(\Omega) &= \hat{Q}_0(\Omega) +R_{QF}(\Omega) \hat{x}(\Omega) \;. \label{eq33}
\end{align}

To be more specific, let us consider the probe realized by a coherent light field with power $I_0$ and frequency $\omega_0$. According to the derivation in \cite{PRLsupplement}, 
the $Q_0$ and $F_0$ are proportional to the phase and amplitude quadratures of the optical field. Then we have the canonical response function $R_{QF}(\Omega) = 1$. Since we do not consider a feedback depending on the position of a test mass, we have $\hat{F}=\hat{F}_0$. The power spectra are
\begin{equation}
S_{Q_0}(\Omega) = \frac{\hbar c^2}{8 I_0 \omega_0} \;, \quad S_{F_0}(\Omega) = \frac{8I_0 \hbar \omega_0}{c^2} \;, \nonumber
\end{equation}
and the cross-correlation vanishes, $S_{F_0Q_0}(\Omega) =0$.

Since $R_{QF}(\Omega) = 1$ for our optical probe, using Eqs.~(\ref{eq27})-(\ref{eq33}), we have the error and disturbance in the Fourier space
\begin{align}
\hat{N}(\Omega) &= \hat{Q}_0(\Omega) +R_{xx}(\Omega) \hat{F}_0(\Omega) \;, \nonumber \\
\hat{D}(\Omega) &= R_{px}(\Omega) \hat{F}_0(\Omega) \;. \nonumber
\end{align}
Thus the power spectra of the error and disturbance are
\begin{align}
S_{\epsilon} &= S_{Q_0}+ \left| R_{xx} \right|^2 S_{F_0} + 2 {\rm{Re}} \left[ R_{xx} S_{F_0Q_0} \right] \;, \label{eq11} \\
S_{\eta} &=  \left|R_{px} \right|^2 S_{F_0} \label{eq12a} \;. 
\end{align}

\begin{figure*}[t]
\centering
\includegraphics[width=170mm]{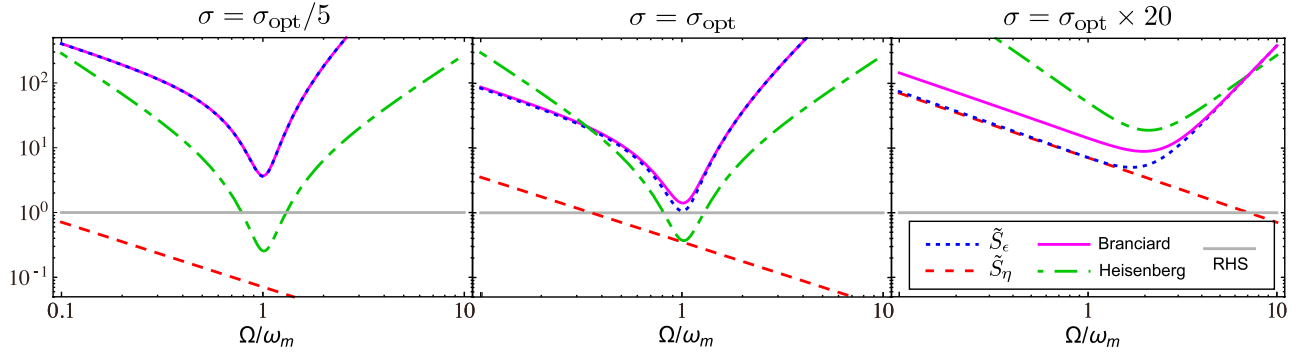}
\caption{EDR applied to a damped harmonic oscillator. For the Branciard inequality, each curve represents $\tilde{S}_{\epsilon}$ (blue, dotted), $\tilde{S}_{\eta}$ (red, dashed), and the LHS of Eq.~(\ref{eq83}) (magenta, solid), respectively. For the Heisenberg inequality, the LHS of Eq.~(\ref{eq101}) (green, dashed-dotted) is plotted. The horizontal line is the right-hand side (RHS) of the inequalities, $|\tilde{\chi}(\Omega)|^2=1$. From the left to the right, $\sigma \equiv I_0 \omega_0 /(mc^2 \omega_m^2)=\sigma_{\rm{opt}}/5$, $\sigma_{\rm{opt}}$, and $\sigma_{\rm{opt}}\times20$, respectively. $\kappa$ is fixed to $\rho \equiv \kappa/\omega_m=0.3$.}
\label{fig4}
\end{figure*}

We model the test mass as a harmonic oscillator coupled with environmental thermal bath. Free evolution of the position and momentum of the damped oscillator without thermal contributions are given by, e.g. \cite{Khalili:2012vj},
\begin{align}
\hat{x}_0(t) &= e^{-\kappa_m t/2} \left[ \hat{x}_{\rm{ini}} \cos \omega_mt +\frac{\hat{p}_{\rm{ini}}}{m \omega_m} \sin \omega_mt \right] \;, \nonumber \\
\hat{p}_0(t) &=e^{-\kappa_m t/2} \left[ - m \omega_m \hat{x}_{\rm{ini}} \sin \omega_mt + \hat{p}_{\rm{ini}} \cos \omega_m t \right] \nonumber \\
&-\frac{m \kappa_m}{2} \hat{x}_0(t) \;, \nonumber
\end{align}
where $m$, $\kappa_m$, and $\omega_m$ are the mass, the decay rate, and the resonant frequency of the oscillator. The initial values are $\hat{x}_{\rm{ini}} =\hat{x}_0(0)$ and $\hat{p}_{\rm{ini}} =\hat{p}_0(0)$. The commutator decays as $e^{-\kappa_m t/2}$ but it is compensated by thermal fluctuations from the external heat bath.  The two-time commutation relation between the position operators at different times, $t$ and $t^{\prime}$, are computed straightforwardly and 
then from Eq.~(\ref{eq50}), the response functions in the Fourier space are obtained,
\begin{align}
R_{xx}(\Omega) &= \frac{1}{2m \omega_m} \left\{ \frac{1}{\Omega + \omega_m+i\kappa_m /2 } - \frac{1}{\Omega -\omega_m+i\kappa_m /2}  \right\}  \;, 
\label{eq21} \\
R_{xp}(\Omega) &= - i \frac{W}{4\omega_m} \left\{ \frac{e^{-i \theta}}{\Omega + \omega_m+i\kappa_m /2 } + \frac{e^{i \theta}}{\Omega -\omega_m+i\kappa_m /2}  \right\}  \;, \nonumber \\
R_{px}(\Omega) &= -R_{xp}(\Omega) \;, \nonumber 
\end{align}
where $W\equiv \sqrt{4\omega_m^2 + \kappa_m^2}$ and $\theta \equiv \arctan [-\kappa_m/(2\omega_m)]$. Then using the formula \cite{PRLsupplement}
\begin{equation}
[ \hat{X}_0(\Omega),\hat{Y}_0^{\dag}(\Omega^{\prime}) ] = -2\pi i \hbar \delta (\Omega-\Omega^{\prime}) \left\{ R_{XY}(\Omega) - R_{YX}^{*}(\Omega^{\prime}) \right\} \;, \label{eq100} 
\end{equation}
and comparing with Eq.~(\ref{eq66}), we have
\begin{equation}
\chi(\Omega) = \frac{2 \kappa_m \Omega^2}{\{ (\Omega+\omega_m)^2+ \kappa_m^2/4 \} \{ (\Omega-\omega_m)^2+ \kappa_m^2/4 \}} \;. 
\label{eq7}
\end{equation}

The power spectrum of test-mass zero-point fluctuations are given by the fluctuation-dissipation theorem, $S_{x_0}=2 \hbar\, {\rm{Im}}[R_{xx}(\Omega)]$. Using Eq.~(\ref{eq21}),
\begin{align}
S_{x_0} (\Omega) &= \frac{2 \hbar\, \kappa_m \Omega}{m\{ (\Omega+\omega_m)^2 +\kappa_m^2/4 \}\{ (\Omega-\omega_m)^2 +\kappa_m^2/4 \}} \;, 
\label{eq8} \\
S_{p_0} (\Omega) &= m^2 \Omega^2 S_{x_0}(\Omega) \;. \label{eq9a}
\end{align}
Since $\langle x_{\rm{ini}} \rangle=0$ and $\langle p_{\rm{ini}} \rangle=0$, we have $S_{\sigma_A} = S_{x_0}$ and $S_{\sigma_B} = S_{p_0}$. 

First we see that the Robertson's inequality in Eq.~(\ref{eq6}) is always saturated for a damped harmonic oscillator as explicitly verified with Eqs.~(\ref{eq7})--(\ref{eq9a}). Consequently, for such a minimum uncertainty state, the third term in the LHS of Eq.~(\ref{eq82}) vanishes, and the inequality, generally keeping nonzero correlation term $S_{F_0Q_0}$, reads
\begin{equation}
\tilde S_\epsilon+\tilde S_\eta= \frac{S_{Q_0} + 2 |R_{xx}|^2 S_{F_0} +2\,\mathrm{Re}[R_{xx}S_{F_0Q_0}]}{ 2\hbar \,{\mathrm{Im}} [R_{xx}]} \ge 1\,. 
\label{eq24}
\end{equation}
If $S_{F_0Q_0}$ remains real-valued, using the Braginsky's inequality $
S_{Q_0} S_{F_0} -|S_{F_0Q_0}|^2 \ge \hbar^2$ and after some algebra (derived in \cite{PRLsupplement}), we obtain
\begin{equation}
\tilde S_\epsilon+\tilde S_\eta \ge \sqrt{2+(2-\sqrt{2})(a^2+b^2)+(3-2\sqrt{2})a^2b^2} \;, \label{eq22}
\end{equation}
where $a \equiv S_{F_0Q_0}/\hbar$ and $b \equiv \mathrm{Re}[R_{xx}]/\mathrm{Im}[R_{xx}]$.
While it is possible for $S_{F_0Q_0}$ to be complex, this usually leads to the modification of the test mass's dynamics, which we shall not consider. From this equation, we see that the best one can do is to have $a=b=0$, that is, $S_{F_0Q_0}=0$ and $\mathrm{Re} [R_{xx}]=0$, which takes place on resonance, where the inequality is a factor of $\sqrt{2}$ from saturation.

In Fig.~\ref{fig4}, we show the normalized error, disturbance, LHS of the Branciard EDR in Eq.~(\ref{eq83}), and LHS of the Heisenberg inequality in Eq.~(\ref{eq101}). There are three characteristic regimes of the inequality, depending on the dimensionless parameter $\sigma \equiv I_0 \omega_0 /(mc^2 \omega_m^2)$, which determines the intensity of a laser power or the strength of disturbance. At smaller $\sigma$, $\tilde{S}_{\epsilon}$ dominates the inequality, while at larger $\sigma$, both $\tilde{S}_{\epsilon}$ and $\tilde{S}_{\eta}$ increase in the same way. Since $\tilde{S}_{\epsilon}$ and $\tilde{S}_{\eta}$ are in the tradeoff relation for middle $\sigma$, there exits the optimal $\sigma$ to minimise the LHS of the Branciard inequality at a certain frequency. One can show that the LHS is the closest to the saturation at the resonant frequency. The optimal disturbance strength there is given by
$\sigma_{\rm{opt}}= \rho\sqrt{1+\rho^2/16}/(8\sqrt{2})$
where $\rho \equiv \kappa/\omega_m$. With the optimal disturbance, the LHS of the inequality gives $\sqrt{2}$ so that the inequality always holds but is not saturated. While the Heisenberg inequality is violated at around the resonant frequency in the small $\sigma$ regime. This is understood from the inequality obtained by using the HUR, $\tilde S_\epsilon+\tilde S_\eta \geq 2 \sqrt{\tilde S_\epsilon \tilde S_\eta} \geq 2 \left| \tilde{\chi}(\Omega) \right|$. For a damped harmonic oscillator, the lower bound is $2$, while the minimum of LHS is $\tilde S_\epsilon+\tilde S_\eta=\sqrt{2}$, which clearly violates the HUR. 

The gap $\sqrt{2}$ from the saturation in the Branciard EDR arises from the fact that we have a factor of 2 in front of the $S_{F_0}$ term in Eq.~(\ref{eq24}). This is because of the disturbing force of the probe, $S_{F_0}$, included in both the error and disturbance in Eqs.~(\ref{eq11}) and (\ref{eq12a}). Since there are two tradeoff relations between $S_{Q_0}$ and $S_{F_0}$ (standard quantum limit \cite{Braginsky:1974SvPhU,Caves:1980rv}) in the quantity $\tilde{S}_{\epsilon}+\tilde{S}_{\eta}$, both contribute to the lowest value of the LHS. One might think that the correlation between the probe variables $\hat{Q}_0$ and $\hat{F}_0$, for example, squeezed light, helps reach the saturation. However, this does not work as we show above. It would be interesting to study whether this deviation from the saturation is specific to a damped harmonic oscillator or universal by applying to other systems.

{\it{Conclusions}} --- We have extended a universally valid EDRs proposed for a discrete measurement to a continuous measurement and derived the EDRs in the Fourier space in terms of the power spectra of the system and probe variables. Application to a linear optomechanical system, particularly to a damped harmonic oscillator, clearly shows the tradeoff relation between the error and the disturbance and leads to the existence of the optimal strength of the disturbance. The EDR in the Fourier space would be experimentally testable with, for instance, an optomechanical system in which zero-point fluctuations of an system contribute and would give a new insight into understanding of quantum foundations.

\begin{acknowledgments}
The authors would like to thanks to Yiqiu Ma and Huan Yang for valuable discussions. A.N. is supported by JSPS Postdoctoral Fellowships for Research Abroad. Y.C.\ is supported by NSF grant PHY-1404569 and CAREER Grant PHY-0956189.
\end{acknowledgments}

\bibliography{/Volumes/USB-MEMORY/my-research/bibliography}

\clearpage

\newpage

\appendix

\section{Supplemental material for "Universally Valid Error-Disturbance Relations in Continuous Measurements"}

\subsection{Definitions of power spectra}

For an arbitrary observable $\hat{X}$, the Fourier transform is defined by
\begin{equation}
\hat{X}(t) \equiv \int_{-\infty}^{\infty} \frac{d\Omega}{2\pi} e^{-i \Omega t} \hat{X}(\Omega)\;. \label{eq17}
\end{equation}
Then the one-sided power spectral densities are defined by 
\begin{align}
\langle \hat{X}(\Omega) \hat{X}^{\dag}(\Omega^{\prime})+ \hat{X}^{\dag}(\Omega^{\prime})\hat{X}(\Omega) \rangle \equiv 2\pi \delta(\Omega-\Omega^{\prime}) S_{X} (\Omega) \;, 
\label{eq60} \\
\langle \hat{X}(\Omega) \hat{Y}^{\dag}(\Omega^{\prime})+ \hat{Y}^{\dag}(\Omega^{\prime})\hat{X}(\Omega) \rangle \equiv 2\pi \delta(\Omega-\Omega^{\prime}) S_{XY} (\Omega) \;, \nonumber
\end{align}
which satisty the properties
\begin{align}
S_{X} (\Omega) &=S_{X}^{*} (\Omega)=S_{X} (-\Omega) \;, \nonumber \\
S_{YX} (\Omega) &=S_{XY}^{*} (\Omega)=S_{XY} (-\Omega) \;. \nonumber
\end{align}
If a system is stationary, the correlation function depends only on the difference of times $t$ and $t^{\prime}$. Denoting it by $\tau \equiv t-t^{\prime}$, the correlation functions are related to the power spectral densities by the Wiener-Khintchine theorem:
\begin{align}
\frac{1}{2}\langle \hat{X}(\tau)\hat{X}(0)+\hat{X}(0)\hat{X}(\tau) \rangle &= \frac{1}{2} \int_{-\infty}^{\infty} \frac{d\Omega}{2\pi} e^{-i\Omega \tau} S_{X} (\Omega) \;, \label{eq10} \\
\frac{1}{2}\langle \hat{X}(\tau)\hat{Y}(0)+\hat{Y}(0)\hat{X}(\tau) \rangle &= \frac{1}{2} \int_{-\infty}^{\infty} \frac{d\Omega}{2\pi} e^{-i\Omega \tau} S_{XY} (\Omega) \;. \nonumber
\end{align}
In particular, taking $\tau \rightarrow 0$ in Eq.~(\ref{eq10}), we obtain the autocorrelation function 
\begin{equation}
\langle \hat{X}^2(t) \rangle = \frac{1}{2} \int_{-\infty}^{\infty} \frac{d\Omega}{2\pi} S_{X} (\Omega) = \int_{0}^{\infty} \frac{d\Omega}{2\pi} S_{X} (\Omega) \;. \nonumber
\end{equation}

\subsection{Useful formulas}
We derive convenient formulas in the spectral domain.

{\bf{Formula 1:}} If $\hat{X}(t)$ is an Hermitian operator, then 
\begin{equation} 
[ \hat{X}(\Omega), \hat{X}^{\dag}(\Omega) ]   =0 \;.
\label{eq23}
\end{equation}

\noindent \underline{proof}
\begin{equation}
\hat{X}^2(t) = \int \frac{d\Omega}{2\pi} \int \frac{d\Omega^{\prime}}{2\pi}\, e^{-i(\Omega +\Omega^{\prime}) t}   \hat{X}(\Omega)\hat{X}(\Omega^{\prime})   \;. \nonumber  
\end{equation}
Integrating the above equation in time gives 
\begin{align}
\int dt\, \hat{X}^2(t)
&= \int \frac{d\Omega}{2\pi} \int \frac{d\Omega^{\prime}}{2\pi}\,   \hat{X}(\Omega)\hat{X} (\Omega^{\prime})  2\pi \delta(\Omega +\Omega^{\prime}) \nonumber \\
&= \int \frac{d\Omega}{2\pi} \,   \hat{X}(\Omega)\hat{X} (-\Omega)   \nonumber \\
&= \int \frac{d\Omega}{2\pi} \,   \hat{X}(\Omega)\hat{X}^{\dag} (\Omega)   \;. \label{eq15}
\end{align}
At the last line, we used 
\begin{equation}
\hat{X}^{\dag}(\Omega) =\hat{X}(-\Omega) \;, \label{eq18}
\end{equation}
which is derived from Eq.~(\ref{eq17}) for an Hermitian operator. If we represent Eq.~(\ref{eq15}) as the integral with respect to $\Omega^{\prime}$, the expression becomes
\begin{equation}
\int \frac{d\Omega^{\prime}}{2\pi} \,   \hat{X}^{\dag}(\Omega^{\prime})\hat{X} (\Omega^{\prime})   \;. \nonumber
\end{equation}
Since the variable $X$ is an arbitrary at each frequency, we obtain
\begin{equation}
  \hat{X}(\Omega)\hat{X}^{\dag} (\Omega)   =   \hat{X}^{\dag} (\Omega)\hat{X}(\Omega) \;, \nonumber
\end{equation} 
that is,
\begin{equation}
  [ \hat{X}(\Omega), \hat{X}^{\dag}(\Omega) ]   =0 \;. \nonumber
\end{equation}
\hfill $\blacksquare$\\

{\bf{Formula 2:}} For an arbitrary Hermitian operators, $\hat{X}(t)$ and $\hat{Y}(t)$,
they satisfy the following relation in the Fourier domain,
\begin{equation}
[ \hat{X}_0(\Omega),\hat{Y}_0^{\dag}(\Omega^{\prime}) ] = -2\pi i \hbar \delta (\Omega-\Omega^{\prime}) \left\{ R_{XY}(\Omega) - R_{YX}^{*}(\Omega^{\prime}) \right\} \;, \nonumber
\end{equation}
with the response function defined in Eq.~(\ref{eq34}). \\

\noindent \underline{proof}
\begin{align}
&[ \hat{X}_0(\Omega),\hat{Y}_0(\Omega^{\prime}) ] \nonumber \\
&=\int_{-\infty}^{\infty} dt\, \int_{-\infty}^{\infty} dt^{\prime}\, e^{i\Omega t} e^{i\Omega^{\prime} t^{\prime}} [ \hat{X}_0(t),\hat{Y}_0(t^{\prime}) ] \nonumber \\
&=\left(   \int_{-\infty}^{\infty} dt \, \int_{-\infty}^{\infty} dt^{\prime}\, \Theta (t-t^{\prime}) + \int_{-\infty}^{\infty} dt^{\prime} \, \int_{-\infty}^{\infty} dt \, \Theta (t^{\prime}-t)  \right) \nonumber \\
& \quad \times e^{i\Omega t} e^{i\Omega^{\prime} t^{\prime}} [ \hat{X}_0(t),\hat{Y}_0(t^{\prime}) ] \nonumber \\
&= -i\hbar \left\{ \int_{-\infty}^{\infty} dt^{\prime}\, e^{i(\Omega +\Omega^{\prime}) t^{\prime}} R_{XY}(\Omega) \right. \nonumber \\
& \quad \quad \;\; \left. - \int_{-\infty}^{\infty} dt \, e^{i(\Omega+\Omega^{\prime}) t} R_{YX}(\Omega^{\prime}) \right\} \nonumber \\
&= -2\pi i \hbar \delta(\Omega+\Omega^{\prime}) \left\{ R_{XY}(\Omega)-R_{YX}(\Omega^{\prime}) \right\} \;. \nonumber 
\end{align}
Changing $\Omega^{\prime} \rightarrow -\Omega^{\prime}$ and using Eqs.~(\ref{eq18}) and (\ref{eq99}), we have
\begin{equation}
[ \hat{X}_0(\Omega),\hat{Y}_0^{\dag}(\Omega^{\prime}) ] = -2\pi i \hbar \delta (\Omega-\Omega^{\prime}) \left\{ R_{XY}(\Omega) - R_{YX}^{*}(\Omega^{\prime}) \right\} \;. \nonumber
\end{equation}
\hfill $\blacksquare$

\subsection{Derivation of error-disturbance relations in continuous measurements}

We denote the system's observables by $\hat{A}$ and $\hat{B}$ and suppose that the system's variable $\hat{A}$ is measured by a meter variable of a probe $\hat{M}$. Following the Ozawa's definitions in a discrete measurement \cite{Ozawa:2003PhLA,Ozawa:2003PhRvA}, we define two operators $\hat{N}$ and $\hat{D}$ in a continuous measurement: 
\begin{align}
\hat{N} (t) &\equiv \hat{M} (t) - \hat{A}_0 (t) \;, \label{eq87} \\
\hat{D} (t) &\equiv \hat{B} (t) - \hat{B}_0 (t) \;, \label{eq88}  
\end{align}
where the subscript stands for free evolution. Then the error and disterbance induced by the measurements are 
\begin{align}
\epsilon_A^2 (t) &\equiv \langle \hat{N}^2 (t) \rangle = \langle \{ \hat{M}(t) - \hat{A}_0(t) \}^2 \rangle \;, 
\label{eq85}  \\
\eta_B^2 (t) &\equiv \langle \hat{D}^2 (t) \rangle = \langle \{ \hat{B}(t) - \hat{B}_0(t) \}^2 \rangle \;. 
\label{eq84}
\end{align}
Here $\langle \cdots \rangle$ denotes the ensemble average. The variances of observables $\hat{A}_0$ and $\hat{B}_0$ are
\begin{align}
\sigma_A^2 (t) &\equiv \langle \{ \hat{A}_0(t)- \langle \hat{A}_0(t) \rangle \}^2 \rangle = \langle \hat{A}_0^2(t) \rangle - \langle \hat{A}_0(t) \rangle^2 \;, \nonumber \\
\sigma_B^2 (t) &\equiv \langle \{ \hat{B}_0(t)- \langle \hat{B}_0(t) \rangle \}^2 \rangle = \langle \hat{B}_0^2(t) \rangle - \langle \hat{B}_0(t) \rangle^2 \;. \nonumber
\end{align}

Following the same proof of the universally valid error-disturbance relation by Ozawa in a discrete measurement \cite{Ozawa:2003PhLA,Ozawa:2003PhRvA}, we show that the following error-disturbance relation holds in a continuous measurement:
\begin{equation}
\epsilon_A(t) \eta_B(t) + \epsilon_A(t) \sigma_{B}(t) + \sigma_{A}(t) \eta_B(t) \geq | C_{AB}(t,t) | \;,
\label{eq9} 
\end{equation}
where
\begin{equation}
C_{AB}(t,t^{\prime}) \equiv \frac{i}{2} \langle [ \hat{A}_0(t), \hat{B}_0(t^{\prime}) ] \rangle \;. \label{eq51}
\end{equation}

\noindent \underline{proof}

Since $\hat{M}(t)$ and $\hat{B}(t)$ reside in different Hilbert spaces, then $[\hat{M}(t),\hat{B}(t)]=0$ gives
\begin{align}
&[\hat{N}(t),\hat{D}(t)] +[\hat{N}(t),\hat{B}_0(t)] +[\hat{A}_0(t),\hat{D}(t)] \nonumber \\
& \quad =-[\hat{A}_0(t),\hat{B}_0(t)]\;. \nonumber
\end{align}
Taking the expectation values and using the triangle inequality, we have
\begin{align}
& \left| \langle [\hat{N}(t),\hat{D}(t)] \rangle \right| + \left| \langle [\hat{N}(t),\hat{B}_0(t)] \rangle \right| + \left| \langle [\hat{A}_0(t),\hat{D}(t)] \rangle \right| \nonumber \\
& \quad \geq \left| \langle [\hat{A}_0(t),\hat{B}_0(t)] \rangle \right| \;.
\label{eq26}
\end{align}
From the definitions of the error and disturbance in Eqs.~(\ref{eq85}) and (\ref{eq84}), the expextation value of a squared operator is always larger than its variance. Then
\begin{equation}
\epsilon_A(t)\geq \sigma_N(t) \;, \quad \quad \eta_B(t) \geq \sigma_D (t) \;. 
\label{eq25}
\end{equation}
Using the Robertson's inequality \cite{Robertson:1929zz},
\begin{equation}
\sigma_{X}(t) \sigma_{Y}(t) \geq \frac{1}{2} \left| \langle [ \hat{X}_0(t), \hat{Y}_0(t) ] \rangle \right| \;, \nonumber 
\end{equation}
which holds for an arbitrary pair of observables, we have
\begin{equation}
\epsilon_A(t) \eta_B(t) \geq \frac{1}{2} \left| \langle [ \hat{N}(t), \hat{D}(t) ] \rangle \right| \;. \nonumber 
\end{equation}
Substituting this for the first term in Eq.~(\ref{eq26}) and using the Robertson's inequality and Eq.~(\ref{eq25}) again for the second and third terms in Eq.~(\ref{eq26}), we obtain
\begin{align}
&\epsilon_A(t) \eta_B(t) + \epsilon_A(t) \sigma_{B}(t) + \sigma_{A}(t) \eta_B(t) \nonumber \\
& \quad \geq \frac{1}{2} | \langle [ \hat{A}_0(t), \hat{B}_0(t) ]\rangle | \;. \nonumber 
\end{align}
\hfill $\blacksquare$\\

In a continuous measurement, it is convenient to work in the Fourier domain. To derive the error-disturbance relation in the Fourier domain, we assume that the whole system is stationary. For the later use, here we introduce a filter function $\Gamma(t)$ and redefine the error and disturbance in Eqs.~(\ref{eq87}) and (\ref{eq88}) as
\begin{align}
\hat{{\cal{N}}} (t) &\equiv \int_{-\infty}^{t} dt^{\prime} \Gamma(t-t^{\prime}) \hat{N}(t^{\prime}) = \hat{{\cal{M}}} (t) - \hat{{\cal{A}}}_0 (t) \;, \nonumber \\
\hat{{\cal{D}}} (t) &\equiv \int_{-\infty}^{t} dt^{\prime} \Gamma(t-t^{\prime}) \hat{D}(t^{\prime}) =\hat{{\cal{B}}} (t) - \hat{{\cal{B}}}_0 (t) \;, \nonumber
\end{align}
where $\hat{{\cal{A}}}_0 (t)$, $\hat{{\cal{B}}}_0 (t)$, $\hat{{\cal{M}}} (t)$, and $\hat{{\cal{B}}} (t)$ are defined in a similar fashion using the same filter.
The new error can be computed in the same way as the nonfiltered case,
\begin{align}
\epsilon_{\cal{A}}^2 & \equiv \langle \hat{{\cal{N}}}^2(t) \rangle \nonumber \\
&= \int_{-\infty}^{t} dt^{\prime} \int_{-\infty}^{t} dt^{\prime\prime} \Gamma(t-t^{\prime}) \Gamma(t-t^{\prime\prime}) \langle \hat{N}(t^{\prime}) \hat{N}(t^{\prime\prime}) \rangle \nonumber \\
&= \frac{1}{2} \int_{-\infty}^{\infty} \frac{d\Omega}{2\pi} \left| \Gamma (\Omega) \right|^2 S_{\epsilon} (\Omega) \;, \nonumber
\end{align}
where we defined
\begin{equation}
\Gamma(\Omega) \equiv \int_{-\infty}^{\infty} d\tau\, \Theta(\tau) \Gamma(\tau)  e^{i\Omega \tau} = \int_{0}^{\infty} d\tau\,\Gamma(\tau)  e^{i\Omega \tau} \;, \nonumber
\end{equation}
and used Eqs.~(\ref{eq60}) and (\ref{eq23}). As well, $\eta_{\cal{B}}$, $\sigma_{{\cal{A}}_0}$, and $\sigma_{{\cal{B}}_0}$ are given by
\begin{align}
\eta_{\cal{B}}^2 &= \frac{1}{2} \int_{-\infty}^{\infty} \frac{d\Omega}{2\pi} \left| \Gamma (\Omega) \right|^2 S_{\eta} (\Omega) \;, \nonumber \\
\sigma_{\cal{A}}^2 &= \frac{1}{2} \int_{-\infty}^{\infty} \frac{d\Omega}{2\pi} \left| \Gamma (\Omega) \right|^2 S_{\sigma_{A}} (\Omega) \;, \nonumber \\
\sigma_{\cal{B}}^2 &= \frac{1}{2} \int_{-\infty}^{\infty} \frac{d\Omega}{2\pi} \left| \Gamma (\Omega) \right|^2 S_{\sigma_{B}} (\Omega) \;. \nonumber
\end{align}

For the filtered quantities, the Ozawa's inequality in Eq.~(\ref{eq9}) is replaced with
\begin{equation}
\epsilon_{\cal{A}} \eta_{\cal{B}} + \epsilon_{\cal{A}} \sigma_{\cal{B}} + \sigma_{\cal{A}} \eta_{\cal{B}} \geq | C_{\cal{A}\cal{B}} | \;.
\label{eq14}
\end{equation}
Note that the time dependence in the original inequality disappears because we are assuming a stationary system. The right-hand side is  
\begin{align}
| C_{\cal{A}\cal{B}} | &= \frac{1}{2} \left| \langle [ \hat{\cal{A}}_0(t), \hat{\cal{B}}_0(t)] \rangle \right| \nonumber \\
&= \frac{1}{2} \left| \int_{-\infty}^{t} dt^{\prime} \int_{-\infty}^{t} dt^{\prime\prime} \Gamma(t-t^{\prime}) \Gamma(t-t^{\prime\prime}) \right. \nonumber \\
&\times \left. \langle [\hat{A}_0(t^{\prime}), \hat{B}_0(t^{\prime\prime})] \rangle \right| \nonumber \\
&= \frac{1}{2} \left| \int_{-\infty}^{\infty} \frac{d\Omega^{\prime}}{2\pi} \int_{-\infty}^{\infty} \frac{d\Omega^{\prime\prime}}{2\pi} \langle [ \hat{A}_0(\Omega^{\prime}), \hat{B}_0^{\dagger}(\Omega^{\prime\prime})] \rangle \right. \nonumber \\
& \times \left. e^{-i (\Omega^{\prime}-\Omega^{\prime\prime}) t} \Gamma(\Omega^{\prime})\Gamma^{\ast}(\Omega^{\prime\prime}) \right| \;.
\label{eq28} 
\end{align}
If the system observables satisfy the commutation relation with the following form 
\begin{equation}
[ \hat{A}_0(\Omega), \hat{B}^{\dagger}_0(\Omega^{\prime})] = 2\pi i \hbar \delta (\Omega-\Omega^{\prime}) \hat{\chi}(\Omega) \;, 
\label{eq102}
\end{equation}
Eq.~(\ref{eq28}) gives
\begin{align}
| C_{\cal{A}\cal{B}}| = \frac{\hbar}{2} \left| \int_{-\infty}^{\infty} \frac{d\Omega^{\prime}}{2\pi} \langle \hat{\chi}(\Omega^{\prime}) \rangle |\Gamma(\Omega^{\prime})|^2 \right| \;. \nonumber
\end{align}
Substituting the above quantities for Eq.~(\ref{eq14}), we can obtain the Ozawa's inequality in the Fourier space. However, the expression includes many integrals and is difficult to be dealt with. Since the band-pass filter we introduced can be chosen arbitrarily, we apply a narrow band-pass filter so as to extract the contribution from a certain frequency range centered at $\Omega = \Omega^{\prime}, \Omega^{\prime \prime}$ with its width $\Delta \Omega$. Then the both sides of Eq.~(\ref{eq14}) gives the same factor $|\Gamma(\Omega)|^2 \Delta \Omega /(2\pi)$, which are canceled out. Therefore, we finally obtain
\begin{align}
&\sqrt{S_{\epsilon} (\Omega) S_{\eta} (\Omega)} + \sqrt{S_{\epsilon} (\Omega) S_{\sigma_B} (\Omega)} + \sqrt{S_{\sigma_A} (\Omega) S_{\eta} (\Omega)} \nonumber \\
& \quad \geq \hbar \left| \langle \hat{\chi}(\Omega) \rangle \right| \;. \nonumber
\end{align}

This procedure of the derivation can also be applied to other inequalities. We do not repeat the derivation, but we can express the Branciard's inequality \cite{Branciard:2013PNAS,Branciard:2014PhRvA} and the Robertson's inequality \cite{Robertson:1929zz} in terms of power spectra as well. As a result, for a continuous measurement we obtain in the Fourier space the Branciard inequality 
\begin{align}
&S_{\epsilon} (\Omega)S_{\sigma_B} (\Omega) + S_{\sigma_A} (\Omega)S_{\eta} (\Omega) \nonumber \\
& \quad +2 \sqrt{S_{\epsilon} (\Omega)S_{\eta} (\Omega) \left\{ S_{\sigma_A} (\Omega)S_{\sigma_B} (\Omega) - \hbar^2 \left| \langle \hat{\chi}(\Omega) \rangle \right|^2 \right\}} \nonumber \\
& \quad \geq \hbar^2 \left| \langle \hat{\chi}(\Omega) \rangle \right|^2 \;. \nonumber
\end{align}
and the Robertson's inequality
\begin{equation}
\sqrt{S_{\sigma_A}(\Omega)S_{\sigma_B} (\Omega)} \geq \hbar \left| \langle \hat{\chi}(\Omega) \rangle \right| \;. \nonumber
\end{equation}

\subsection{A linear system}

If a system has Hamiltonian that is at most quadratic in its canonical coordinates and momenta, the system is called a linear system. Any linear combination of the canonical coordinates and momenta and a complex number (c-number) defines a linear observable of the system. In the linear system, it is shown that in the Heisenberg picture, the commutator of the operators of any two linear observables at two times gives a c-number \cite{Buonanno:2001kr}. In what follows, we summalize the basic propaties of a linear system.

Suppose that $\hat{A}$ is an arbitrary observable of a system and that $\hat{M}$ is an arbitrary observable of a probe. We assume that the interaction between the probe and the system is the von Neumann type, whose Hamiltonian in the interaction picture is given by
\begin{equation}
\hat{H}_I(t) = - \hat{x} _0(t) \hat{F}_0(t) \;, \nonumber
\end{equation}
where $\hat{x} _0(t)$ is the observable of the system to be measured and $\hat{F}_0(t)$ is the generalized force of the probe. Since the variables in different Hilbert spaces always commute, the commutation relations between $\hat{A}_0$ ($\hat{M}_0$), which is the freely evolving part of $\hat{A}$ ($\hat{M}$), and the interaction Hamiltonian become
\begin{align}
[\hat{H}_I(t^{\prime}), \hat{A} _0(t)] &= -2i C_{xA}(t^{\prime},t) \hat{F}_{0}(t^{\prime}) \;, \label{eq30} \\
[\hat{H}_I(t^{\prime}), \hat{M} _0(t)] &= 2i C_{FM}(t^{\prime},t) \hat{x} _0(t^{\prime})  \;. 
\label{eq32}
\end{align}
Here the correlation functions for freely evolving variables are that defined in Eq.~(\ref{eq51}). The time evolution of an arbitrary linear observable $\hat{X}(t)$ in the Heisenberg picture is given by, e.g.~\cite{Buonanno:2001kr},
\begin{widetext}
\begin{equation}
\hat{X}(t) = \hat{X}_0(t) + \frac{i}{\hbar} \int_{-\infty}^{t} dt^{\prime} [ \hat{H}_I (t^{\prime}), \hat{A}_0(t) ] + \left( \frac{i}{\hbar} \right)^2 \int_{-\infty}^{t} dt^{\prime} \int_{-\infty}^{t^{\prime}} dt^{\prime\prime} [ \hat{H}_I (t^{\prime\prime}),[ \hat{H}_I (t^{\prime}), \hat{X}_0(t) ] ]+ \cdots \;, 
\label{eq31}
\end{equation}
\end{widetext}
where $\hat{X}_0(t)$ is a freely evolving observable.

In Eq.~(\ref{eq31}), replacing $\hat{X}(t)$ with $\hat{A}(t)$ and $\hat{M}(t)$ and using Eqs.~(\ref{eq30}) and (\ref{eq32}) interatively, we obtain the time evolution of $\hat{A}$ and $\hat{M}$
\begin{align}
\hat{A}(t) &= \hat{A}_0(t) +\frac{2}{\hbar} \int_{-\infty}^{t} dt^{\prime} C_{Ax}(t,t^{\prime}) \hat{F}(t^{\prime})  \;, \nonumber \\
\hat{M}(t) &= \hat{M}_0(t) + \frac{2}{\hbar} \int_{-\infty}^{t} dt^{\prime} C_{MF}(t,t^{\prime}) \hat{x}(t^{\prime})  \;. \nonumber 
\end{align}
where the first terms are the variables in the free evolution and the second terms are contribution added by measurement procedures. 

When the above formulas are applied to a linear optomechanical system, in which the position of a system is measured by a optical probe, we identify the position and momentum of the test mass as $\hat{A}=\hat{x},\hat{p}$ and the position of the probe as $\hat{M}=\hat{Q}$. Then the dynamic of the system and the probe is governed by a set of the following equations:
\begin{align}
\hat{x}(t) &= \hat{x}_0(t) + \frac{2}{\hbar} \int_{-\infty}^{t} dt^{\prime} C_{xx}(t,t^{\prime}) \hat{F}(t^{\prime})\;, 
\label{eq68} \\
\hat{p}(t) &= \hat{p}_0(t) + \frac{2}{\hbar} \int_{-\infty}^{t} dt^{\prime} C_{px}(t,t^{\prime}) \hat{F}(t^{\prime})\;, \\
\hat{Q}(t) &= \hat{Q}_0(t) + \frac{2}{\hbar} \int_{-\infty}^{t} dt^{\prime} C_{QF}(t,t^{\prime}) \hat{x}(t^{\prime}) \;. 
\label{eq69}
\end{align}

The Fourier transforms of Eqs.~(\ref{eq68})-(\ref{eq69}) are
\begin{align}
\hat{x}(\Omega) &= \hat{x}_0(\Omega) + R_{xx}(\Omega) \hat{F}(\Omega)  \;, \nonumber \\
\hat{p}(\Omega) &= \hat{p}_0(\Omega) + R_{px}(\Omega) \hat{F}(\Omega) \;,  \nonumber \\
\hat{Q}(\Omega) &= \hat{Q}_0(\Omega) +R_{QF}(\Omega) \hat{x}(\Omega) \;, \nonumber
\end{align}
where we defined
\begin{align}
R_{XY}(\Omega) &\equiv \frac{2}{\hbar} \int_{-\infty}^{\infty} d\tau\, \Theta(\tau) C_{XY}(\tau)  e^{i\Omega \tau} \nonumber \\
&= \frac{2}{\hbar} \int_{0}^{\infty} d\tau\,C_{XY}(\tau)  e^{i\Omega \tau} \;. 
\label{eq34}
\end{align}
From the definition of $R_{XY}(\Omega)$, we have
\begin{equation}
R^{*}_{XY}(\Omega) = R_{XY}(-\Omega) \;. \label{eq99}
\end{equation}

\subsection{Optical probe} 

Suppose that carrier light with the classical amplitude $E_0$ and the frequency $\omega_0$ is injected to measure the position of an object. According to the two-photon formalism in \cite{Kimble:2000gu}, an electromagnetic field of the carrier light including vacuum fluctuations is written as
\begin{align}
E_a(t)&= [E_0+E_{a1}(t)]\; \cos \omega_0 t +E_{a2}(t)\; \sin \omega_0 t 
\label{eq11a} \\ 
E_{a1,2}(t) &= \sqrt{\frac{4\pi \hbar \omega_0}{{\cal{A}}c}} \int_{0}^{\infty}(a_{1,2}\;e^{-i\Omega t}+a_{1,2}^{\dag} \;e^{i\Omega t}) \frac{d\Omega}{2\pi} \;, 
\label{eq12}
\end{align}
with quadrature annihilation operators
\begin{equation}
a_1 \equiv \frac{a_{+}+a_{-}^{\dag}}{\sqrt{2}} \;, \quad \quad a_2 \equiv \frac{a_{+}-a_{-}^{\dag}}{\sqrt{2}i} \;, \nonumber
\end{equation}
where ${\cal{A}}$ is an effective beam area, $\omega_0$ is the carrier frequency, and $c$ is the speed of light. The annihilation operators, $a_{\pm} \equiv a(\omega_0 \pm \Omega)$, satisfy the commutation relations
\begin{align}
[a_{\pm}(\Omega),a_{\pm}^{\dag}(\Omega^{\prime}) ] &= 2 \pi \delta (\Omega-\Omega^{\prime})\;, \nonumber \\
[a_{\pm}(\Omega),a_{\pm}(\Omega^{\prime}) ] &= 0 \;, \nonumber
\end{align} 
then the quadrature annihilation operators satisfy
\begin{align}
&[a_1(\Omega),a_2^{\dag}(\Omega^{\prime}) ]=-[a_2(\Omega),a_1^{\dag}(\Omega^{\prime}) ]= 2 i \pi \delta (\Omega-\Omega^{\prime})\;, \nonumber \\
&[a_1(\Omega),a_1(\Omega^{\prime}) ]=[a_1(\Omega),a_1^{\dag}(\Omega^{\prime}) ]=[a_1(\Omega),a_2(\Omega^{\prime}) ]= 0 \;. \nonumber
\end{align}

We suppose that light is reflected at a test mass, e.g. a tiny mirror whose size is neglected. We denote the small displacement of the mirror by $x=c \Delta t/2$ (here we treat the mirror dispalcement as a c-number only for the purpose to derive the relevant quantities of the light probe). Writing an input field as $a$ and a output field as $b$ and substituting Eqs.~(\ref{eq11a}) and (\ref{eq12}) into the relation $E_{b}(t)=E_{a}(t -\Delta t )$ with the assumption that the sideband frequency is much smaller than that of a carrier, that is, $\omega_0 \gg \Omega$, we have the input-output relation for the quadrature operators at the leading order in the small quantities $\Omega/\omega$, $x$ and $a_{1,2}$,
\begin{align}
b_{1} &= a_{1} \;, \nonumber \\
b_{2} &= a_{2} + \sqrt{\frac{8 I_0 \omega_0}{\hbar c^2 }} x \;,
\label{eq65}
\end{align}
where the DC component of the injected laser power is
\begin{equation}
I_0= \frac{E_0^2 {\cal{A}}c}{8\pi} \;. \nonumber
\end{equation}	 
From Eq.~(\ref{eq65}), we identify the position operator for the electromagnetic field as
\begin{equation}
\hat{Q}_0(\Omega) = \sqrt{\frac{\hbar c^2}{8 I_0 {\omega_0}}} a_2(\Omega) \;.
\label{eq61}
\end{equation}

Next let us calculate radiation pressure exerted on the test mass. The radiation pressure is given by
\begin{equation}
\hat{F}_{\rm{rp}}(t)=\frac{2 I(t)}{c}=\frac{{\cal{A}}}{2\pi} E_{a}^2(t) \;, \nonumber
\end{equation}
where $I$ is an optical power injected to the test mass. The factor 2 comes from that the light is reflected by the test mass. By smoothing out the rapidly oscillating terms with the frequency $\omega_0$, the fluctuating component with the period of a sideband is
\begin{align}
\Delta \hat{F}_{\rm{rp}} (t) &= \frac{{\cal{A}}}{2\pi} E_0 E_{a1}(t) \nonumber \\
&= \sqrt{\frac{8 I_0 \hbar \omega_0}{c^2}} \;\int_{-\infty}^{\infty} a_1 e^{-i\Omega t} \frac{d\Omega}{2\pi} \;. \nonumber
\end{align}
Thus we define the stochastic force exerted by the measurement
\begin{equation}
\hat{F}_0(\Omega) = \sqrt{\frac{8 I_0 \hbar \omega_0}{c^2}}  a_1(\Omega) \;. 
\label{eq62}
\end{equation}

From Eqs.~(\ref{eq61}) and (\ref{eq62}), we have
\begin{align}
[ \hat{Q}_0(\Omega), \hat{F}_0^{\dag}(\Omega^{\prime}) ] = - 2\pi i \hbar \delta (\Omega -\Omega^{\prime}) \;, \label{eq72} \\
[ \hat{Q}_0(\Omega), \hat{Q}_0^{\dag}(\Omega^{\prime}) ] =[ \hat{F}_0(\Omega), \hat{F}_0^{\dag}(\Omega^{\prime}) ] = 0 \;. \nonumber
\end{align}
Since the Fourier transform of Eq.~(\ref{eq72}) is 
\begin{equation}
[ \hat{Q}_0(t), \hat{F}_0(t^{\prime}) ] = - i \hbar \delta(t-t^{\prime}) \;, \nonumber
\end{equation}
by writing $\tau=t-t^{\prime}$, the response function in the Fourier space is
\begin{align}
R_{QF}(\Omega) &\equiv \frac{2}{\hbar} \int_{-\infty}^{\infty} d\tau \Theta(\tau) C_{QF}(\tau) e^{i \Omega \tau}  \nonumber \\
&= 1 \;. 
\end{align}

Using Eq.~(\ref{eq60}) and assuming the sidebands of the probe is in a coherent vacuum state (no correlation between $\hat{Q}$ and $\hat{F}$), namely,
\begin{equation}
\langle a_i(\Omega)a_j^{\dag}(\Omega^{\prime}) + a_j^{\dag}(\Omega^{\prime}) a_i(\Omega) \rangle = 2 \pi \delta (\Omega-\Omega^{\prime}) \delta_{ij} \;, \nonumber
\end{equation} 
we obtain the power spectra for the coherent light probe
\begin{equation}
S_{Q_0}(\Omega) = \frac{\hbar c^2}{8 I_0 \omega_0} \;, \quad S_{F_0}(\Omega) = \frac{8I_0 \hbar \omega_0}{c^2} \;, \quad S_{F_0Q_0}(\Omega) =0 \;. \nonumber
\end{equation}

\subsection{Braginsky's inequality}
\label{sec:Braginsky}

In a linear system, all commutation relation gives a c-number. Suppose that the Hermitian variables $\hat{A}$ and $\hat{B}$ of the probe satisfy the commutation relations:
\begin{equation}
[ \hat{A}(\Omega), \hat{B}^{\dag}(\Omega^{\prime}) ] = 2\pi i \hbar\, \chi (\Omega) \delta (\Omega-\Omega^{\prime}) \;. \label{eq22e} 
\end{equation}
The Braginsky's inequality can be derived from the commutation relation in Eqs.~(\ref{eq22e}) and the Schwarz's inequality for anbitrary quantum states, $|\alpha \rangle$ and $|\beta \rangle$:
\begin{equation}
\langle \alpha |\alpha \rangle \langle \beta |\beta \rangle \geq | \langle \alpha | \beta \rangle |^2 \;. \nonumber
\end{equation}

Let us consider the case with $|\alpha \rangle= \hat{A}(\Omega) |\cdot \rangle $ and $|\beta \rangle= \hat{B}(\Omega)|\cdot \rangle$, where $|\cdot \rangle$ denotes an arbitrary state. The Schwarz's inequality gives
\begin{equation}
\langle \cdot | \hat{A}^{\dag}(\Omega) \hat{A}(\Omega) | \cdot \rangle \langle \cdot | \hat{B}^{\dag}(\Omega) \hat{B}(\Omega) | \cdot \rangle \geq | \langle \cdot | \hat{A}^{\dag}(\Omega) \hat{B}(\Omega) | \cdot \rangle |^2 \;.
\label{eq24a}
\end{equation}
From Eq.~(\ref{eq23}), the Hermite-conjugate operators in the Fourier space commute, 
\begin{align}
\langle \cdot | \hat{A}^{\dag}(\Omega) \hat{A}(\Omega) | \cdot \rangle &= \frac{1}{2} \langle \cdot | \hat{A}(\Omega) \hat{A}^{\dag}(\Omega) + \hat{A}^{\dag}(\Omega) \hat{A}(\Omega) | \cdot \rangle \nonumber \\
&= \frac{1}{2} 2\pi \delta (0) S_{A}(\Omega) \;, \nonumber \\
\langle \cdot | \hat{B}^{\dag}(\Omega) \hat{B}(\Omega) | \cdot \rangle &= \frac{1}{2} \langle \cdot | \hat{B}(\Omega) \hat{B}^{\dag}(\Omega) + \hat{B}^{\dag}(\Omega) \hat{B}(\Omega) | \cdot \rangle \nonumber \\
&= \frac{1}{2} 2\pi \delta (0) S_{B}(\Omega) \;, \nonumber 
\end{align}
On the other hand,
\begin{align}
& \langle \cdot | \hat{A}^{\dag}(\Omega) \hat{B}(\Omega) | \cdot \rangle \nonumber \\
&= \frac{1}{2} \langle \cdot | \hat{A}^{\dag}(\Omega) \hat{B}(\Omega) + \hat{B}(\Omega) \hat{A}^{\dag}(\Omega) + [ \hat{A}^{\dag}(\Omega), \hat{B}(\Omega) ] | \cdot \rangle \nonumber \\
&= \frac{1}{2} 2\pi \delta (0) S_{BA}(\Omega) +\frac{1}{2} \langle \cdot | [ \hat{A}^{\dag}(\Omega), \hat{B}(\Omega) ] | \cdot \rangle \nonumber \\
&= \frac{1}{2} 2\pi \delta (0) S_{BA}(\Omega) -\frac{1}{2} \langle \cdot | [ \hat{A}(\Omega), \hat{B}^{\dag}(\Omega) ]^{\dag} | \cdot \rangle \nonumber \\
&=  \frac{1}{2} 2\pi \delta (0) \left\{ S_{BA}(\Omega) + i\hbar \chi^{*}(\Omega) \right\} \;, \nonumber
\end{align}
where we used Eqs.~(\ref{eq22e}). Substituting these into Eq.~(\ref{eq24a}), we obtain
\begin{align}
& S_{A}(\Omega)S_{B}(\Omega) \nonumber \\
& \geq \left| S_{BA}(\Omega) + i\hbar \chi^{*}(\Omega) \right|^2 \nonumber \\
&= |S_{BA}(\Omega)|^2 + \hbar^2 |\chi(\Omega)|^2+2 \hbar\, {\rm{Im}} [\chi(\Omega)S_{BA}(\Omega)] \;. \nonumber
\end{align}
Thus, we have
\begin{equation}
S_{A}(\Omega)S_{B}(\Omega) - |S_{BA}(\Omega)|^2 \geq \hbar^2 |\chi(\Omega)|^2+2 \hbar\, {\rm{Im}} [\chi(\Omega) S_{BA}(\Omega)] \;. \label{eq39}
\end{equation}

\subsection{Derivation of Eq.~(\ref{eq22})} 

The LHS of the Branciard inequality in the case of a damped harmonic oscillator is given by 
\begin{equation}
\tilde S_\epsilon+\tilde S_\eta= \frac{S_{Q_0} + 2 |R_{xx}|^2 S_{F_0} +2\,\mathrm{Re}[R_{xx}S_{F_0Q_0}]}{ 2\hbar \,{\mathrm{Im}} [R_{xx}]} \ge 1\,. \nonumber 
\end{equation}
We assume that $S_{F_0Q_0}$ is real-valued. Using the Braginsky's inequality $
S_{Q_0} S_{F_0} -|S_{FQ_0}|^2 \ge \hbar^2$ in Eq.~(\ref{eq39}), we have
\begin{align}
\tilde S_\epsilon+\tilde S_\eta &\geq \frac{2 \sqrt{2 |R_{xx}|^2 S_{Q_0} S_{F_0}} +2\,\mathrm{Re}[R_{xx}S_{F_0Q_0}]}{ 2\hbar \,{\mathrm{Im}}[R_{xx}]} \nonumber \\
&\geq \frac{2 \sqrt{2 |R_{xx}|^2 (\hbar^2 + |S_{F_0Q_0}|^2)} +2\,\mathrm{Re}[R_{xx}S_{F_0Q_0}]}{ 2\hbar \,{\mathrm{Im}} [R_{xx}]} \nonumber \\
&= \sqrt{2(1+a^2)(1+b^2)}+ab \nonumber \\
&\geq \sqrt{2(1+a^2)(1+b^2)}-\sqrt{a^2b^2} \nonumber 
\end{align}
where
\begin{equation}
a \equiv \frac{S_{FQ_0}}{\hbar} \;, \quad b \equiv \frac{{\mathrm{Re}}[R_{xx}]}{{\mathrm{Im}}[R_{xx}]} \;. \nonumber
\end{equation}
Since
\begin{align}
& (\sqrt{2(1+a^2)(1+b^2)}-\sqrt{a^2b^2})^2 \nonumber \\
&= 2(1+a^2)(1+b^2) +a^2b^2 -2 \sqrt{2 a^2b^2 (1+a^2)(1+b^2)} \nonumber \\
&\geq 2(1+a^2)(1+b^2) +a^2b^2 \nonumber \\
&\quad -\sqrt{2} \left\{ a^2(1+b^2) + b^2 (1+a^2) \right\} \nonumber \\
&= 2+(2-\sqrt{2})(a^2+b^2)+(3-2\sqrt{2})a^2b^2 \;, \nonumber 
\end{align}
the LHS of the Branciard inequality is constrained from below
\begin{align}
\tilde S_\epsilon+\tilde S_\eta 
&\geq \sqrt{2+(2-\sqrt{2})(a^2+b^2)+(3-2\sqrt{2})a^2b^2} \nonumber \\
&\geq \sqrt{2} \;. \nonumber
\end{align}

\end{document}